\documentclass[usenatbib]{mn2e}
\usepackage[T1]{fontenc}

\usepackage{graphicx}
\usepackage{float}
\usepackage{arydshln}
\usepackage{verbatim}
\usepackage{amsmath}
\usepackage{color}
\usepackage{aas_macros}

\newcommand{\tblskip}{\noalign{\vskip 1.5mm}}

\newcommand{\titleA}{Combining Spectroscopic and Photometric Surveys}
\newcommand{\titleB}{Same or  different sky?}

\newcommand{\nbright}{71}
\newcommand{\width}{0.5\textwidth}
\newcommand{\xfigure}[1]{
\begin{center}
\includegraphics[width=\width]{links/#1}
\end{center}
}

\newcommand{\citedir}{\citet}
\newcommand{\mycite}{\citep}
\newcommand{\citeind}{\citep}
\newcommand{\citeoverlap}{\citeind{cai2011,gazta,cai2012,kirk,mcdonald,deputter}}
\newcommand{\be}{\begin{equation}}
\newcommand{\ee}{\end{equation}}
\newcommand{\fomc}{\text{FoM}_{w\gamma}}

\newcommand{\xbf}{\begin{figure}}
\newcommand{\xef}{\end{figure}}

\title{Combining Spectroscopic and Photometric Surveys: \\
Same or  different sky?}

\title[\titleA: \titleB]{\titleA: \\ \titleB}

\author[Martin Eriksen, Enrique Gazta\~naga]
{\parbox{\textwidth}{Martin Eriksen$^{1,2}$, Enrique Gazta\~naga$^1$}
\vspace{0.4cm}\\
$^1$Institut de Ci\`encies de l'Espai (IEEC-CSIC),  E-08193 Bellaterra (Barcelona), Spain \\
$^2$Leiden Observatory, Leiden University, PO Box 9513, NL-2300 RA Leiden, Netherlands \\
}

\newcommand{\gkone}{FxB-$\left<\delta_F \gamma_F\right>$}
\newcommand{\gktwo}{FxB-$\left<\delta_B \gamma_F\right>$}
\newcommand{\gkthree}{FxB-$\left<\delta \gamma \right>$}
\newcommand{\gkfb}{FxB-$\left<\text{FB}\right>$}
\newcommand{\fb}{$\left<\text{FB}\right>$}

\begin{document}
\maketitle
\begin{abstract}
This article looks at the combined  constraints from a photometric and
spectroscopic survey. These surveys will measure cosmology using weak lensing
(WL), galaxy clustering, baryon acoustic oscillations (BAO) and redshift space
distortions (RSD). We find, contrary to some findings in the recent literature,
that overlapping surveys can give important benefits when measuring dark
energy. We therefore try to clarify the status of this issue with a full
forecast of two stage-IV surveys using a new approach to properly account for
covariance between the different probes in the overlapping samples. The benefit of
the overlapping survey can be traced back to two factors: additional
observables and  sample variance cancellation. Both needs to be taken into
account and contribute equally when combining 3D power spectrum and 2D
correlations for lensing. With an analytic example we also illustrate that
for optimal constraints, one should minimize the (Pearson) correlation
coefficient between  cosmological and nuisance parameters and maximize the one
among nuisance parameters (e.g. galaxy bias) in the two samples. This can be
achieved by increasing the overlap between the spectroscopic and photometric
surveys. We show how BAO, WL and RSD contribute to this benefit and also look at
some other survey designs, such as photometric redshift errors and
spectroscopic density.
\end{abstract}

\section{Introduction}
The characteristics of a  galaxy survey are in practice limited by time
constraints. One can spend the time going deeper or overlapping different
probes, or going wider.  The issue of overlapping surveys has been an open
question in the literature for the last few years. Several groups have been
working on the topic, trying to find out if it is better to combine
spectroscopic and photometric galaxy surveys on the same or different parts of
the sky. Besides the benefit of finding  an optimal design, these studies can
also help in understanding the best way to combine different probes.

This paper proceeds a series of related papers by the same authors on this
topic. The first deals with modeling of the correlation function. The second
studies the relative impact of WL, RSD and BAO on cosmological forecast. The
third focus on the impact of galaxy bias. From now, we will refer to these as
paper-I\citeind{paperI}, paper-II\citeind{paperII} and
paper-III\citeind{paperIII}. In this paper we focus on the combined
constraints from a photometric and spectroscopic survey.

The combination of spectroscopic and weak lensing surveys helps
to reduce the statistical errors on cosmological parameters  \citeoverlap.
To a great extend the reduction comes from the complementarity (and independence)
of the probes used in WL, RSD and BAO. That is something
which can be done when combining two surveys over separate parts of the sky. In
addition, overlapping surveys includes additional
cross-correlations between the two galaxy samples which could in principle
add or reduce the above benefits.

The galaxy density fluctuations  follow the underlying dark matter
fluctuations. This is something we can actually observe. There is also good
agreement between the shape of the dark matter power spectrum and the galaxy
power spectrum, or the corresponding 2D-correlations.  These are related with a
conversion factor called the galaxy bias $b$. The bias factor depends on how
well the galaxies are tracing the underlying mass. This is again dependent on
the galaxy types and magnitudes. Therefore, splitting or selecting galaxies
will give samples with different characteristics. 

These different galaxy
probes are not independent because they trace the same underlying
matter. However this can result in sample variance cancellation and reduce the
errors on the cosmological parameters. A multiple tracer technique is a method
already suggested in the literature to reduce sampling variance within a survey
\cite{mcdonseljak,cai2012,asorey2}. In this article, the sampling variance can
also cancel between the photometric and spectroscopic survey, between WL and
RSD or galaxy counts of different galaxy types combination, as long as we are
careful to include the full covariance between then. The relative impact of
this cancellation will strongly depend on how much correlated are these probes
relative to how much information there is in each separate measurement.

The first section in this paper introduces the forecast assumptions.
These are already detailed in the other papers of these series, therefore the
explanation is kept to a minimum. The second section presents the forecast of
different survey configurations and probes, studying how these contribute to
the benefit of overlapping galaxy surveys. In section three we present a
generic analytical example to help interpreting the sample variance
cancellation in the forecast. In the fourth section, we study how the galaxy
density and redshift uncertainty affect the conclusions on overlapping surveys.
We end with a conclusion.

\section{Forecast assumptions}
\label{sec:assumpt}
In this section we briefly present the forecast assumptions, which corresponds
directly to the setup in a series of papers (paper I, II, III). While the
description here provide the most essential assumptions, the reader is referred
to paper-II for more detailed information and discussion. A study of
2D-correlations in narrow bins can be found in paper-I.

\paragraph*{Observables and Fisher forecast.}
The forecast study the combined constraints on dark energy and simple
deviations from general relativity from intrinsic galaxy clustering, RSD and
WL. A spectroscopic survey with excellent redshift information is
ideal to measure RSD and intrinsic galaxy clustering, while a photometric
survey allow for shape measurements to probe weak lensing. To simplify the survey
combination, we analyze both surveys using 2D-correlations of galaxy count
overdensities and galaxy shear. See paper-I for a detailed treatment.

We use the Fisher matrix formalism to propagate the covariance of observed
correlations to the covariance (and errors) of the cosmological parameters. Let
$C_x$ be a 2D cross-correlation, with the index $x\equiv\{\ell,z_1,z_2,p_1,
p_2\}$ being a combination of angular multipole scale $\ell$, the two redshift
bins ($z_1, z_2$) and galaxy populations ($p_1, p_2$). The Fisher matrix is
then

\be
F_{\mu \nu} = \sum_{x,y} \frac{\partial C_x}{\partial p_{\mu}} 
  \left(Cov^{-1}\right)_{x,y}
  \frac{\partial C_y}{\partial p_{\nu}}
\label{fisher_matrix}
\ee

\noindent
where $Cov^{-1}$ is the inverse covariance between the observable and 
$\partial C_x / \partial p_{\mu}$ is the derivative with respect to
a parameter $\mu$. The double sum $(x,y)$ is over all considered observables.
Inverting the Fisher matrix,

\be
Cov_{\mu \nu} = {(F^{-1})}_{\mu \nu}
\ee

\noindent
estimate the covariance of the parameters. While the Fisher matrix use a
Gaussian approximation for the parameter likelihood and these errors are lower
bounds, the Fisher matrices are a standard tool for cosmological forecast and
can provide good physical insight.

\paragraph*{Galaxy samples}
\newcommand{\lt}{\lessthan}

\begin{table}
\begin{center}
\begin{tabular}{lrr}
\hline
Parameter & Photometric (F) & Spectroscopic (B)\\
\hline
\noalign{\vskip 0.7mm}
Area [${\text{deg}}^2$] & 14,000  & 14,000 \\
Magnitude limit & $i_{AB}$ < 24.1 & $i_{AB}$ < 22.5 \\
Redshift range & 0.1 < z < 1.5 & 0.1 < z < 1.25 \\
Redshift uncertainty & 0.05(1+z) & 0.001 (1+z) \\
z Bin width; \# bins  & 0.07 (1+z); 12 bins & 0.01(1+z); \nbright\ bins \\
\tblskip
\hdashline
\tblskip
Bias: b(z) & 1.2 + 0.4(z - 0.5)  & 2 + 2(z - 0.5) \\
Shape noise & 0.2 & No shapes \\
\tblskip
\hdashline
\tblskip
density [gal/arcmin$^2$] & 6.5 &  0.4\\
nz - $z_0$ & 0.702 & 0.467 \\
nz - $\alpha$ & 1.274 & 1.913 \\
nz - $\beta$ & 2.628 & 1.083 \\
\hline
\end{tabular}
\end{center}
\caption{Parameters describing the two surveys/populations. The first block give the
area, magnitude limit, redshift range used in the forecast, redshift uncertainty modeled as 
a Gaussian and the redshift bin width. Second block give the galaxy bias ($\delta_g 
= b \delta_m$ and average galaxy shape uncertainty. Third block give the galaxy density
and parameters for the n(z) shape.}
\label{tbl_assumpt}
\end{table}

The forecast in 2D-correlations models the photometric (F for Faint) and
spectroscopic (B for Bright) surveys as two galaxy populations. Forecasting the
parameter errors depends on e.g. the redshift uncertainty, galaxy density and
galaxy bias of both populations. For simplicity, the most important values are
summarized in Table \ref{tbl_assumpt}, while assumptions on minor effects (e.g.
cosmic magnification) and plots can be found in paper-II. The galaxy densities
follow
the distributions

\newcommand{\txtnz}[1]{\frac{dN}{d\Omega dz}}
\newcommand{\nz}[4]{{\left( z/ #2 \right)}^{#3} \exp{\left(-{\left( z / #2 \right)}^{#4}\right)}}

\be
\txtnz{F} \propto \nz{F}{z_0}{\alpha}{\beta}
\ee

\noindent
with the parameters given in the last block of Table \ref{tbl_assumpt}.

Since galaxies often occupy dense regions, the galaxy densities are biased
($\delta_g = b \delta_m$) with respect to the matter distribution. The bias
factor, which depends  both on galaxy formation and selection effects, is an
important uncertainty in the combined forecast. When combining photometric and
spectroscopic surveys, part of the gain comes from improving galaxy bias
constraints. This forecast use one (linear) bias parameter for each redshift
bin and galaxy population. Note that this is different from \cite{gazta}, which
used a smaller number of bias parameters. The fiducial bias evolution $b(z)$
for each population is given in Table \ref{tbl_assumpt}.

\paragraph*{Figure of Merit (FoM)}
A Figure of Merit compress
the ability to measure cosmological parameters
into a single number. While characterizing the full strength of a survey is
more involved, it allows to simple compare the relative strength of surveys,
probes and their combinations. To allow for both measuring the expansion and
growth history, we define

\be
\fomc \equiv \frac{1}{\sqrt{\det\left[\left(F^{-1}\right)_S\right]}}
\label{fomdef}
\ee

\noindent
where $S$ the parameter subspace of $w_0,w_a,\gamma$, and therefore extends the
DETF FoM by including the $\gamma$ parameter, see \citedir{gazta}.  The forecast
include Planck priors (see paper-II) and marginalizing over cosmological
parameters $\Omega_m, \Omega_{DE}, \Omega_{b}, h, \sigma_8, n_s$
\footnote{The fiducial forecast is a $w$CDM model, with values of cosmological
parameters to match the MICE (www.ice.cat/mice) cosmological simulations.  The densities of
matter, dark energy and baryonic matter are respectively $\Omega_m=0.25$,
$\Omega_{DE}=0.75$, $\Omega_b=0.044$. For the dark matter power spectrum P(k),
$\sigma_8=0.8$ is the amplitude of fluctuations in a 8Mpc/h sphere, while $P(k)
= k^{0.95}$ on large scales. The dark energy component has an equation
which equals a cosmological constant. In the forecast, we use the
Eisenstein-Hu power spectrum model, which has the option of removing effects
of BAO.}. We also marginalize over bias parameters, which follows the fiducial
relations in Table  \ref{tbl_assumpt} with one bias parameter in each redshift
bin, separate  for each of the galaxy population. A detailed study of the
galaxy bias can be found in paper-III.

\paragraph*{Non-linear scales}
The forecast only use linear scales. In addition to $\ell_{max} = 300$, which
for technical reasons is used for all correlations, we additionally remove
correlations with too high $k = \ell / \chi(z)$ to prevent low redshift bins to
enter into the non-linear regime. If either redshift bins in a correlations has
$k>k_{max}(z) = \exp(-2.29 + 0.88z)$, the correlation is removed. For further details,
see paper-II.

\section{Benefit of overlapping surveys}
This section present the main arguments of overlapping versus non-overlapping
galaxy surveys. The first subsection present the forecast for different probes
and the contribution of WL, RSD and BAO. The second subsection explains the
contribution from galaxy clustering with multiple tracers and also from
cross-correlations of galaxy counts with background shear. In the third
section, we discuss the sampling variance cancellations that comes from
overlapping surveys tracing the same matter fluctuations. Last subsection
comments on how overlapping surveys also help to reduce systematic effects.

\begin{table*}
\begin{center}
\tabcolsep=0.16cm
\begin{tabular}{lrrrrrc}
\hline
Probe combinations & Fiducial& Fix Bias& No WL& No RSD& No BAO& Observables included \\

\hline
FxB:All&35.5& 213& 7.32& 17.1& 27.8& F+B:All  +  $\left<\delta_B \delta_F\right>$ +  $\left<\delta_B \gamma_F\right>$\\
F+B:All&24.3& 193& 6.13& 11.0& 18.0& F:All + B:All\\
Ratio&1.5& 1.1& 1.2& 1.6& 1.6 & (FxB/F+B):All\\
\tblskip
\hdashline
\tblskip
FxB:Counts&8.3& 61.9& 7.32& 3.31& 6.0 & $\left<\delta_F \delta_F\right>$ + $\left<\delta_B \delta_B\right>$  +  $\left<\delta_B \delta_F\right>$\\
F+B:Counts&6.24& 62.0& 6.13& 1.72& 4.41 & $\left<\delta_F \delta_F\right>$ + $\left<\delta_B \delta_B\right>$\\
Ratio&1.3& 1.00& 1.2& 1.9& 1.4 & (FxB/F+B):Counts\\
\tblskip
\hdashline
\tblskip
F:All&3.10& 51.2& 0.06& 2.61& 2.77& $\left<\delta_F \delta_F\right>$ + $\left<\delta_F \gamma_F\right>$ +  $\left<\gamma_F \gamma_F\right>$\\

B:All&7.9& 52.6& 5.53& 2.76& 5.42& $\left<\delta_B \delta_B\right>$ + $\left<\delta_B \gamma_B\right>$  + $\left<\gamma_B \gamma_B\right>$\\
F:Counts&0.08& 2.85& 0.06& 0.04& 0.05& $\left<\delta_F \delta_F\right>$\\
B:Counts&5.53& 45.4& 5.53& 1.45& 3.78& $\left<\delta_B \delta_B\right>$\\
\tblskip
\hdashline
\tblskip
FxB-$\left<\delta_F \gamma_F\right>$:All&32.4& 201& 7.32& 15.4& 25.3& FxB:All - $\left<\delta_F \gamma_F\right>$\\
FxB-$\left<\delta_B \gamma_F\right>$:All&33.8& 208& 7.32& 16.2& 26.3& FxB:All - $\left<\delta_B \gamma_F\right>$\\
FxB-$\left<\delta \gamma\right>$:All&16.4& 100& 7.32& 7.46& 12.0& FxB:All - $\left<\delta_F \gamma_F\right>$ - $\left<\delta_B \gamma_F\right>$\\
\tblskip
\hdashline
\tblskip
\gkfb:All&30.7& 199& 6.05& 14.1& 23.3 & F+B:All + Cov (same sky)\\
Ratio &1.3& 1.0& 0.99& 1.3& 1.3 & (\gkfb/F+B):All\\
\tblskip
\hdashline
\tblskip
\gkfb:Counts&6.58& 57.1& 6.05& 2.02& 4.67 & F+B:Counts + Cov (same sky)\\
Ratio & 1.1& 0.92& 0.99& 1.2& 1.1 & (\gkfb/F+B):Counts \\
\hline

\end{tabular}
\end{center}
\caption{FoM$_{\gamma w}$ in units of $10^{-3}$. Each row corresponds to a
different combination of probes, including either counts $\delta$ and  shear
$\gamma$ (All) or just galaxy counts (Counts). The combination FxB refers to
the same sky case, while F+B is for separate sky for the photometric (F) and
spectroscopic (B) samples. Block 3 of rows show the single population results.
In block 4,5,6 or rows are the FxB case when removing some of the
cross-correlations as observables, but keep the covariance from overlapping
surveys. This is done for counts-shear ($-\left<\delta \gamma\right>$) or from
all cross-correlations between F and B  samples (-\fb). On the columns is first
the fiducial case, including all the effects. The next columms show the fixed
galaxy bias (i.e. assumed that bias is known), no gravitational lensing
effects, no redshift space distortions effects and no BAO peak. For "No BAO"
the forecast is done without BAO wiggles in the Eisenstein-Hu power spectrum,
while "No RSD" use correlations in real space.}
\label{tbl_fomc}
\end{table*}

\subsection{Increasing FoMs}
Table \ref{tbl_fomc} contains the main forecast table. The two first lines are
respectively for overlapping and non-overlapping surveys, when including both
galaxy counts and shear. The ratio line show the statistical benefit of having
overlapping surveys. For the nominal survey, the $\fomc$ increase with 50\% or
equivalent 30\% in the area. Further, one see that
the result depends on the galaxy bias. Fixing the galaxy bias strongly increase
the $\fomc$, but decrease the benefit of overlapping surveys. For a more detailed
treatment, see paper-III.

Each column either fix the bias (second column) or remove a physical effect,
while keeping the same correlations. Removing effects like doing the forecast in
real space is unobservable, but it is shown to demonstrate the relative effects.
Including WL, RSD and BAO respectively increase $\fomc$ with factors
of 4.8, 2.1 and 1.3. This shows that both lensing and RSD are significant
contributions. Also, while the forecast include the full power spectrum, the
BAO is an important contribution. For a fixed bias or without lensing, the
same-sky ratio drops.  This happens because overlapping surveys better
constrains the free bias parameters and without lensing there are no additional
counts-shear ($\left<\delta \gamma\right>$) cross-correlations. Instructively,
the benefit without BAO and RSD show the competition between higher constraints
in the separate surveys and the benefit of overlapping surveys. The second
section in Table \ref{tbl_fomc} shows the same results using only galaxy
counts (no shear). Results are  qualitative similar to the ones with shear,
but the FoM are smaller, as expected.

To compare the combination of surveys (FxB, F+B) to the constraints to
a single survey, Table 2 include F:All, B:All, F:Counts and B:Counts. The
last two lines show the forecast when only using clustering of galaxy
counts. Despite being deeper, the F sample has much lower $\fomc$ due to larger
photo-z scatter. Constraints from galaxy clustering and RSD is therefore much
higher in spectroscopic than photometric surveys. Also note how removing BAO
(last column) reduce the $\fomc$ by less than half, while removing RSD has a
much larger impact.

Weak lensing also affects the number counts through magnification. The impact is
small for the spectroscopic survey where the RSD dominates. However, for the
photometric (F) sample with only number counts, magnification increase
$\fomc$ by 50\%\footnote{In \citedir{gazta} we used a different and more
confusing notation. There "MAG", which equals what we now label "Counts",
included both magnification and galaxy clustering.}. One should also
note the difference between F:All, B:All and F+B:All. Even if the surveys are
not overlapping, their combined constraints are much higher. This comes from
WL and RSD probing different parameter combinations. Thus, one can greatly
benefit from combining photometric and spectroscopic surveys, even when they do
not overlap.

\subsection{Counts-Shear cross-correlations}
Overlapping surveys allow for the cross-correlation of the two samples. The important
contributions are the cross-correlations of galaxy counts ($\left<\delta_F
\delta_B\right>$), which is a multi tracer approach and the correlations of
spectroscopic galaxy counts with background shear $\left<\delta_B
\gamma_F\right>$. In this subsection we study the effect of counts-shear
correlations ($\left<\delta \gamma\right>$), which either comes from the
cross-correlation of foreground spectroscopic galaxies ($\left< \delta_B,
\gamma_F \right>$) or from within the photometric population ($\left< \delta_F
\gamma_F \right>$).

The count-shear correlations contribute important information. While the
auto-correlations, ignoring RSD, depend on $b^2$, the counts-shear correlations
depend linearly on the bias ($b$). Measuring both the counts-counts
auto-correlations and the count-shear cross-correlations leads to important
improvements. The counts-shear correlations alone give weak bias constraints.
The benefit comes from measuring the galaxy bias with galaxy clustering and then
using these bias measurements to improve cosmological constraints from the
counts-shear cross-correlations.

Section 4 of Table \ref{tbl_fomc} investigates how the different
counts-shear cross-correlations contribute. The three lines corresponds to
FxB:All including different counts-shear correlations. In \gkone, \gktwo\, and
\gkthree, respectively the Faint-Faint, Bright-Faint and all count-shear
cross-correlations are not included. 

Removing either the $\left<\delta_F \gamma_F\right>$ or $\left<\delta_B
\gamma_F\right>$ cross-correlations has less than 7\% effect on the overall
constraints. However dropping both count-shear cross-correlations reduce the
constraints with 52\%. One should be careful and include all counts-shear
cross-correlations in the forecast.  Including counts-shear cross-correlations
for the overlap ($\left<\delta_B \gamma_F\right>$), but ignoring the
information for the photometric population alone ($\left<\delta_F
\gamma_F\right>$), will overestimate the same-sky benefit. We expect this is
the main difference to the findings of \citedir{kirk}.

\subsection{Reduced sample variance}
Besides additional cross-correlations, overlapping volumes directly reduce the
cosmic variance. When two surveys overlap the same volume, the galaxy
overdensities trace the same underlying matter fluctuations. Measuring the same
fluctuations with covariant observables helps in measuring the underlying
matter densities. This multi-tracer technique is a well known method to reduce
the sampling variance \mycite{mcdonseljak,asorey,cai2012}.

Last section in Table \ref{tbl_fomc} quantify the benefit of overlapping
volumes. The forecast of overlapping surveys (FxB) include covariance between
the samples, while the non-overlapping surveys (F+B) are considered
independent. The line \gkfb\ provide a case in-between FxB and F+B, where the
surveys overlap, but without including (as observables) the additional
cross-correlations between the surveys. Thus the only difference between
\gkfb\ and F+B is including the covariance between the samples (B and F) in
\gkfb. Naively one expect a lower $\fomc$ from \gkfb\ than from either FxB or
F+B, since the covariance often reduce the available information.

Including the covariance increase the fiducial $\fomc$ by 26\% (see
(\gkfb/F+B):All), while the effect is only 3\% for a fixed galaxy bias. This
shows the additional covariance from overlapping surveys, increase the $\fomc$
through better bias constraints. For counts alone the increase is smaller.
When bias is known, the additional covariance in the counts reduce, rather than
increase, the FoM. When also including shear, the  bias can also be measured
from $\left<\delta_F \gamma_F\right>$, which explains additional benefits in
(\gkfb/F+B):All than with the Counts alone.

A different forecast method to the 2D-correlations used here, is to combine the
spectroscopic 3D power spectrum and the 2D counts-shear and shear-shear
correlations, see \citedir{gazta}\footnote{In \citedir{gazta} the FxB and F+B
notation did not correspond directly to overlapping and non-overlapping
surveys. Instead F+B was the traditional use of photometric and spectroscopic
surveys, while FxB combined the surveys over the same area and also used the
photometric sample to measure RSD. Combined with problems with the RSD
calculations, in particular underestimating the photo-z effect, this lead to
overestimating the same-sky benefit.}. When combining a 3D and 2D forecast, one
needs to include the covariance between the 2D and 3D estimator. The forecast
will otherwise not properly include sample variance cancellation and be biased
towards a lower same-sky benefit. Not properly including the covariance can
partially explain why \citedir{mcdonald} and \citedir{deputter} find lower
benefit from overlapping surveys. These analysis separate use a 2D correlation
for the transverse modes and a 3D power spectrum for the rest. As the
transverse component accounts for less than 6\% of the modes (according to
their own accounting), they ignore most of the covariance between the galaxy
counts in the photometric and specroscopic sample.

These analysis differs in detail. The parameters included are different, since
our analysis (in this paper) always marginalising over the growth index
($\gamma$), while others marginalise over the neutrino mass, but often
fix the growth. Also the 2D+3D estimation ignore the radial information in the
cross-correlations between the photometric and spectroscopic sample. For the 2D
analysis in narrow bins, this effect give a significant contribution to the
correlations (see paper-I). On the other hand, \citedir{mcdonald} and
\citedir{deputter} use $l_{max}=2000$ for WL (and also larger $k_{max}$ for
BAO), which can reduce the overlap in Fourier space and reduce the same-sky
benefit. We leave modelling the non-linear bias and extending out analysis to
small scales for future work.

\subsection{Control systematics}
This paper focus on the statistical benefit of overlapping surveys. In
addition, and potentially more important, the overlapping surveys provide other
venues for reducing the impact of systematic errors. Examples include measuring
the cross-correlations to reduce the uncertainty in the photometric galaxy
redshift distribution \mycite{newman,newman2,gazta}.

An frequently used approach to handle systematic errors is to parameterize the
unknown quantity, e.g. the galaxy bias. This can remove systematic effects at
the cost of increasing statistical errors. Additional parameters be physically
motivated or selected to follow a mathematical model, e.g. linear in redshift.
In both cases, the cross-correlations can help to constrain these parameters.
One example is the stochastic bias model introduced in paper-III.  When
introducing a stochastic bias parameter, the overlapping surveys are less
affected than non-overlapping surveys. We expect similar behavior might apply
also to other forms of systematics.

\section{A simple example}
\label{sec:example}
Stronger constraints from a higher covariance between observable might seen 
counter intuitive. This section illustrate why covariance can contribute
positively when marginalizing over the galaxy bias.

\subsection{Increased errors}

The covariance matrix for two observables is

\be
Cov = 
\begin{bmatrix}
\sigma_1^2 & k \sigma_1 \sigma_2 \\
k \sigma_1 \sigma_2 & \sigma_2^2
\end{bmatrix}
\ee

\noindent
where $\sigma_1$ and $\sigma_2$ are their errors. The factor $k$ give the cross
correlation (Pearson coefficient) between the observables, with $k=0$ being
independent and $k=1$ fully correlated. Assuming the two observables are equal,
the expected error of x ($\sigma_x$) estimated with a Fisher matrix is

\be
\sigma^2_x = {\left(\frac{\sigma_O}{\partial O/\partial x} \right)}^2
 \left[ \frac{k + 1}{2} \right]
\ee

\noindent
where $\sigma_O$ is the error of the observables. Here uncorrelated ($k=0$)
observables produce the smallest errors, while the limit of fully correlated
observables ($k=1$) is the constraints from one observable alone. If $O_1$ and
$O_2$ are the same observable measured from two surveys, then k is the
fractional overlap in survey areas, ignoring observational noise.

\subsection{General covariance - 3 parameters}

The effect of a covariance changes when introducing nuisance parameters.
Consider the general case of one cosmology parameter (P) and two nuisance
parameters (1, 2). These could be the amplitude of fluctuations ($P=A$) and the
two bias parameters ($b_1,b_2$) from different galaxy samples. Their Fisher
matrix can be written 

\newcommand{\mysep}{\:}
\be
F = 
\begin{bmatrix}
d_P^2    & r_{P 1} \mysep d_P d_1 & r_{P 2} \mysep d_P d_2 \\
r_{P 1} \mysep d_P d_1 &    d_1^2 & r_{1 2} \mysep d_{1} d_{2} \\
r_{P 2} \mysep d_P d_2  & r_{1 2} \mysep d_{1} d_{2}& d_2^2
\end{bmatrix}
\label{Fexample}
\ee

\noindent
where $d_i^2$ is the diagonal component for parameter $i$. The off diagonal
components ($r_{ij} d_i d_j$) are defined so $-1 \leq r_{ij} \leq 1$ are the
Pearson correlation coefficient. For non-overlapping surveys, then $r_{12} = 0$
because observables including the two nuisance parameters are independent.  A
Fisher matrix can estimate the expected parameters variance $\sigma_i^2 =
(F^{-1})_{ii}$ when marginalizing over the uncertainty in the other parameters.
For the parameters $P$, the variance is

\be
\sigma_P^2 = \frac{1}{d^2_P} 
{\left[1 - \frac{r_{P 1}^2 + r_{P 1}^2 - 2 r_{P 1} \, r_{P 2} \, r_{1 2}}{1 - r_{1 2}^2}  \right]}^{-1}
\label{sigmaP}
\ee

\noindent
where the requirement
\be
r^2_{P2} + r^2_{P1} - 2 r_{P1} r_{P2} r_{12} \leq 1 - r_{1 2}^2 
\ee

\noindent
avoid negative values of the variance. From Eq. \ref{sigmaP}, the optimal error
is $\sigma_A = 1/d_P$, which occurs when there is no covariance between P and
the nuisance parameters ($r_{P 1} = r_{P 2} = 0$) . Further, decreasing either
$r_{P1}$ or $r_{P2}$ will lead to better constrains on $P$. For $\alpha \equiv
r_{p1} = r_{p2}$ the expected error on P (Eq.\ref{sigmaP}) is

\be
\sigma_P^2 = \frac{1}{d_P^2} {\left[1 - 
  \left(\frac{2 \alpha^2 }{1+r_{12}} \right)  \right]}^{-1}
 \label{gcovvara}
\ee

\noindent
A larger $r_{12}$ or smaller $\alpha$ reduce the error (and vice versa).  For
optimal constraints, one should minimize the correlation between cosmology and
the nuisance parameters ($\alpha$), while maximizing the correlation between
the nuisance parameters of the two samples ($r_{12}$). We have run some
analytical examples of this and found that, as expected, $r_{12}$ is
proportional to $k$ while alpha depends more weekly on $k$.  This explains
how the error in cosmological parameters could reduce as we increase
the overlap between the two surveys (and therefore $k$).

\subsection{General covariance - n parameters}
The example can be extended to more parameters. Consider the Fisher matrix
where the parameters are divided into the parameters of interest (p) and
nuisance parameters (n). The covariance matrix is the by block inversion 

\begin{align}
F^{-1} &=
{\left[
\begin{array}{cc}
F_{pp} & F_{pn} \\
F_{np} & F_{nn}
\end{array}
\right]}^{-1} \\
&=
\left[
\begin{array}{cc}
S_{nn}^{-1} & - F^{-1}_{pp}  F_{pn} S^{-1}_{pp} \\
- F^{-1}_{nn} F_{np} S^{-1}_{nn} & S_{pp}
\end{array}
\right]
\label{fblockinv}
\end{align}

\noindent
where $F_{xy}$ denote the Fisher matrix subspace for parameter sets $x$ and
$y$ and the Schur complements ($S_{nn}$, $S_{pp}$) are defined by

\begin{align}
\label{shurnn}
S_{nn} &\equiv F_{pp} - F_{pn} F^{-1}_{nn} F_{np} \\
S_{pp} &\equiv F_{nn} - F_{np} F^{-1}_{pp} F_{pn}.
\end{align}.

\noindent
Equivalent to the general $\text{FoM}_{S}$ \citeind{gazta}, for which $\fomc$
(Eq. \ref{fomdef}) is a special case, one have

\begin{equation}
\text{FoM} \equiv \frac{1}{\sqrt{\text{det}\left[(F^{-1})_{pp} \right]}}
= \sqrt{\text{det}\left[F_{pp} - F_{pn} F^{-1}_{nn} F_{np}\right]}
\end{equation}

\noindent
from Eq. \ref{fblockinv} and \ref{shurnn}. The FoM increase with higher
correlation between nuisance parameters (in $F_{nn}$) and lower correlation
between nuisance parameters and parameters of interest ($F_{np}$). Notice
how the cosmological parameters marginalised over is included in the set
of nuisance parameters (n). While the results also depend on the eigenvectors
directions, but we are not discussing this here.

\section{Survey configurations}
\label{sec:config}
The benefit of overlapping surveys depend on the survey specifications.  We
find a same-sky benefit over a wide range of configuration.  However the exact
benefit depend on details and need to be considered when comparing results in
the literature. The combined forecast depends stronger on parameters of the
spectroscopic survey. In this section, we therefore study the effect of
spectroscopic density and the importance of radial information. 

\subsection{Spectroscopic galaxy density}
The number of galaxies is a discrete quantity, which leads to a shot-noise term
in the auto-correlation of galaxy overdensities. A higher spectroscopic density
will reduce the measurement errors and increase the constraints on cosmology.
But for a fixed survey time, there is a  trade off between depth (longer
exposures) and area covered. This subsection ignore survey optimization and
study how increased densities improve constraints for a fixed area.

\xbf
\xfigure{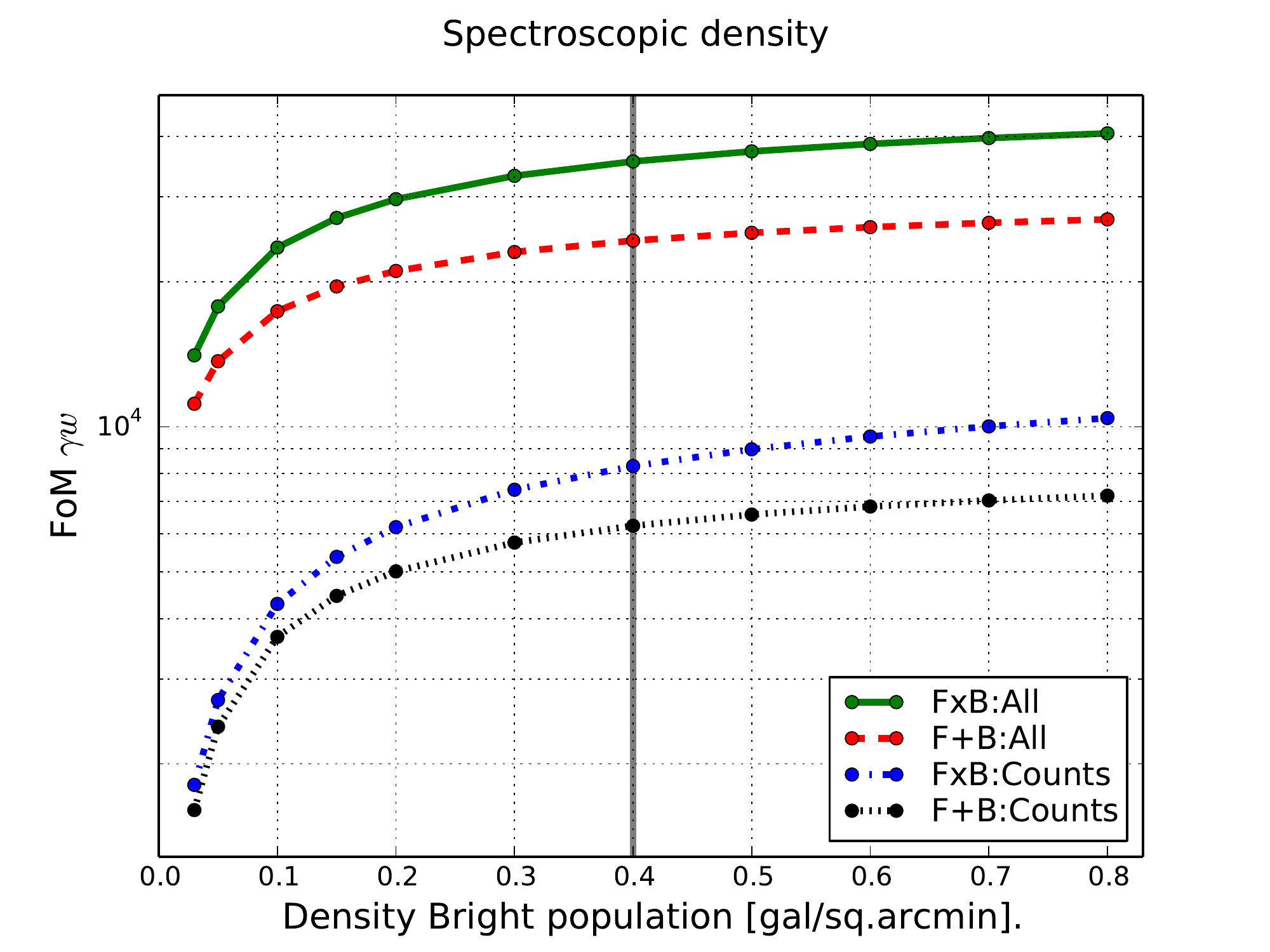}
\xfigure{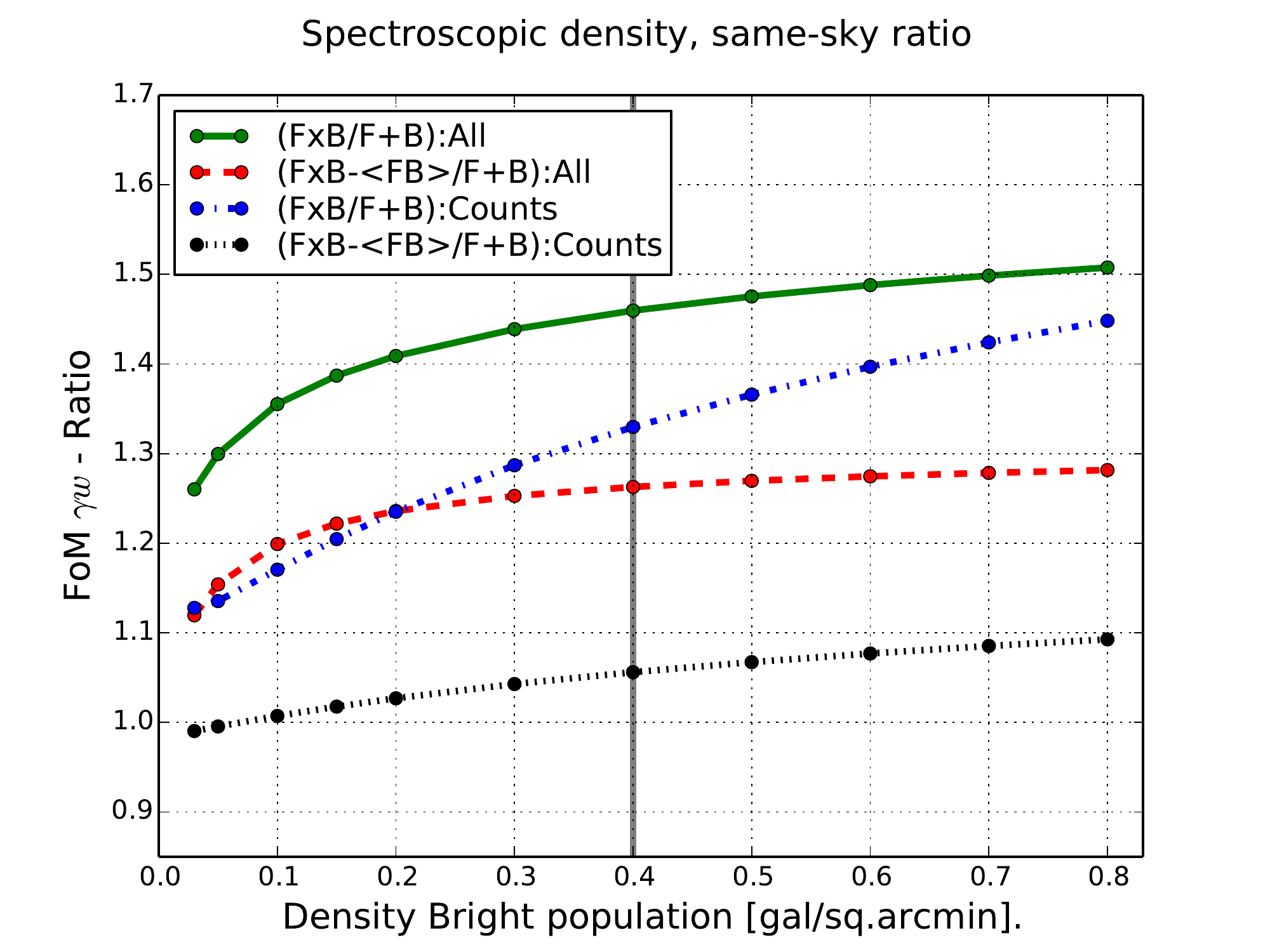}
\caption{Effect of galaxy density in the spectroscopic sample. The top panel
show the absolute $\fomc$, with the galaxy density on the x-axis and the four
lines corresponds to FxB-All, F+B-All, FxB-Counts and F+B-Counts. A vertical
line at 0.4 gal/sq.arcmin marks the fiducial spectroscopic galaxy density. In
the lower panel, the ratios show the same-sky benefit (FxB/F+B) and the effect
of the overlapping volumes (\gkfb/F+B).
}
\label{eff_density}
\xef

Fig. \ref{eff_density} shows the effect of increasing the galaxy density in the
spectroscopic sample. For low densities, the configurations which only include
galaxy counts (FxB-Counts, F+B-Count) drops drastically since the spectroscopic
galaxy clustering dominates the constraints. Naturally, the effect is smaller
when also including WL shear (FxB-All, F+B-All), because the lensing
constraints primarily come from the photometric sample. The lines grow
monotonically and flattens around 0.4 gal/sq.arcmin. Beyond this density, we
only find small improvements by targeting higher densities.  This turnover is
affected by the removal of non-linear scales and the lmax value.

The bottom panel (Fig. \ref{eff_density}) show the same-sky ratio. For low
densities, the error is dominated by shot-noise and sample variance
cancellation becomes less important. Therefore we find that the ratio increase
with density. Also, the (FxB/F+B):Counts ratio grows faster because the
spectroscopic density (B) affects the galaxy counts clustering more than the WL.

\subsection{Redshift uncertainties}
Spectroscopic galaxy surveys typically has excellent redshift determination.
The Gaussian spectroscopic errors we assume ($\sigma_z = 0.001(1+z)$) are 50
times better than the photo-z errors ($\sigma_z = 0.05 (1+z)$). This precision
allows us to measure the radial information in the galaxy clustering.  For this
series of papers, the analysis is done with 2D cross-correlations in narrow
redshift bins. The radial information is encoded in the intrinsic
cross-correlations between the redshift bins (see paper-I, paper-II). This
subsection forecasts the constraints for larger redshift errors for the
spectroscopic (Bright) sample. Although the errors are artificially high, the
results help to understand the benefit of high radial resolution.

\xbf
\xfigure{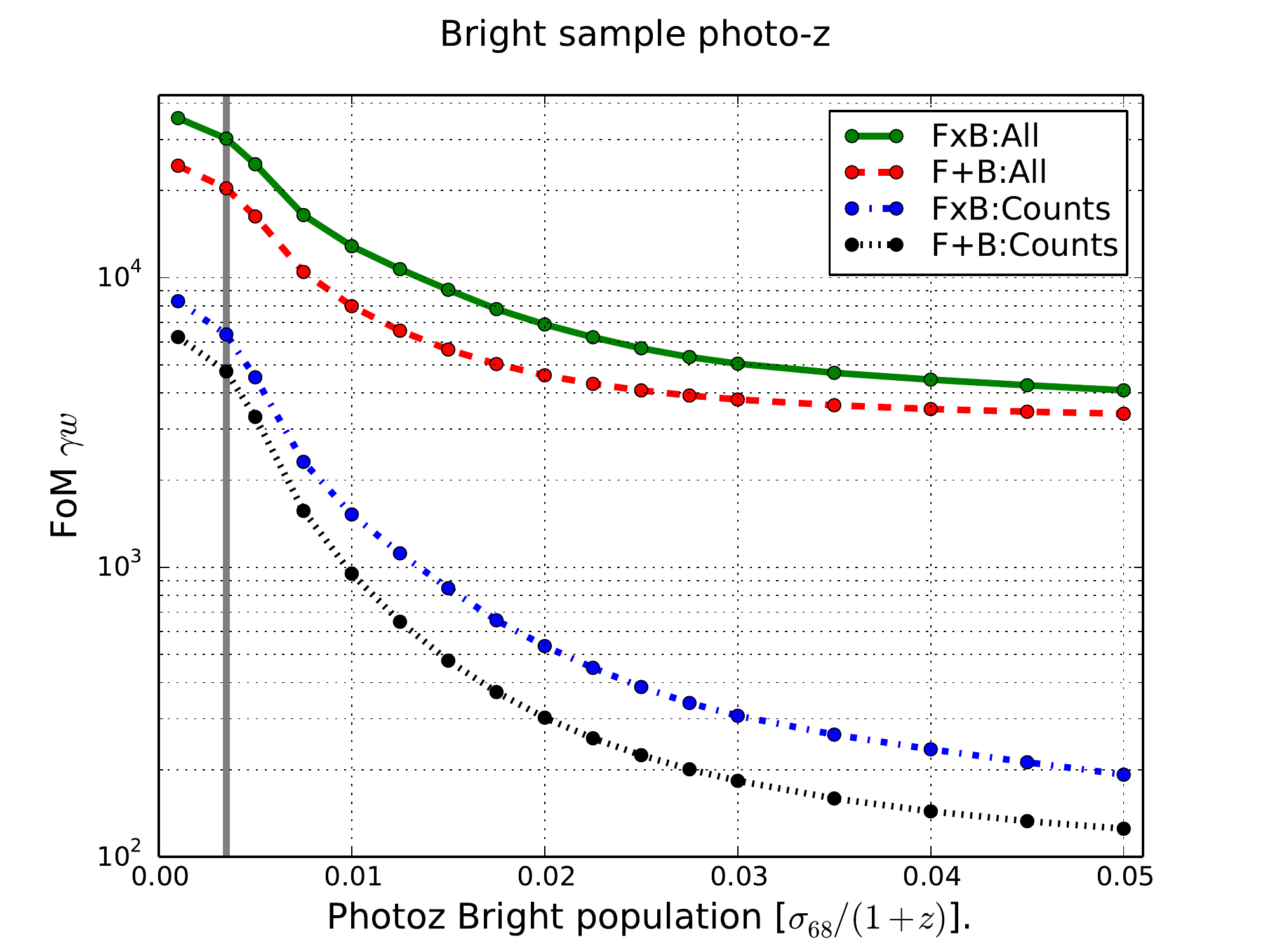}
\xfigure{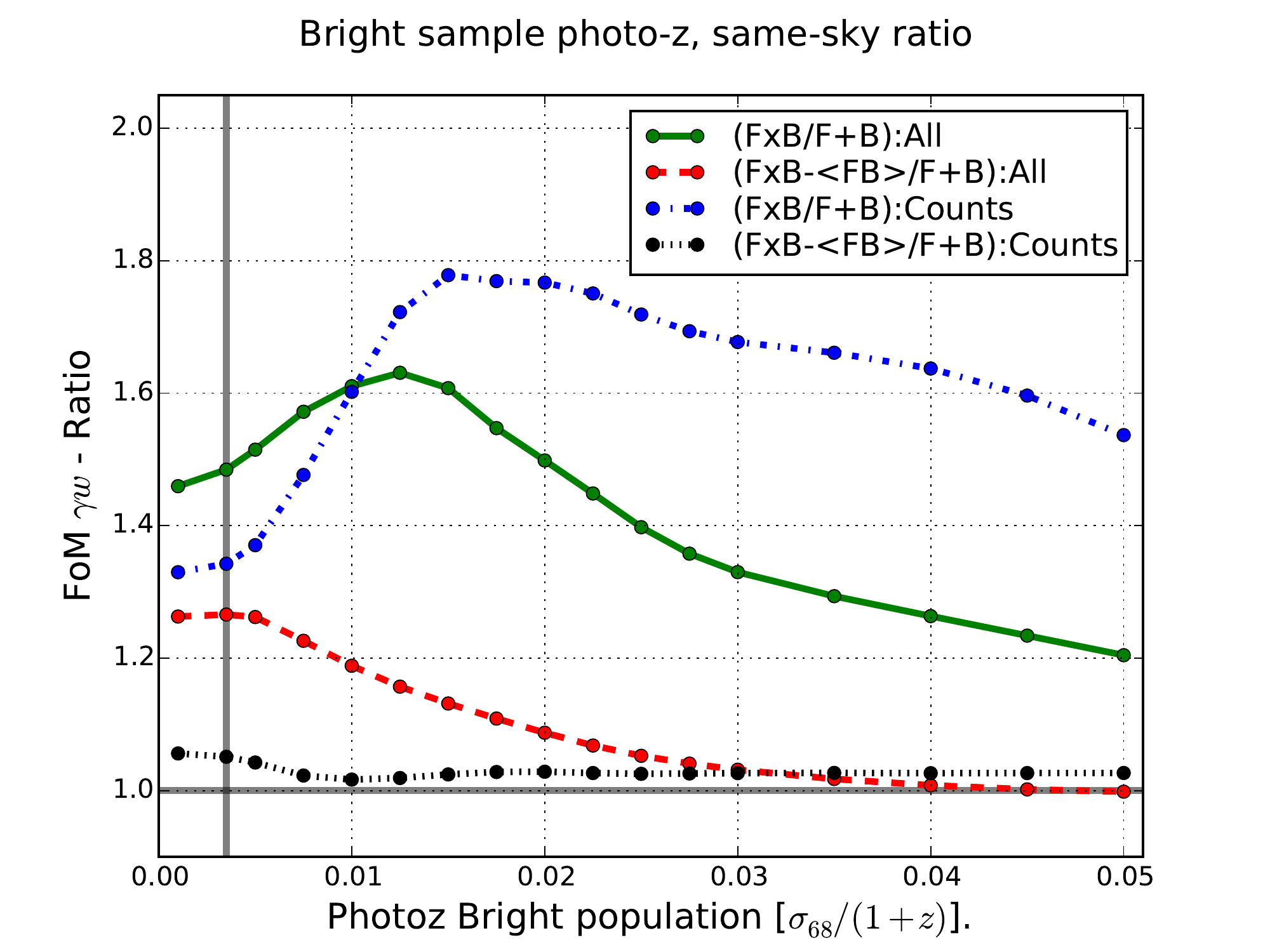}
\caption{Impact of redshift errors of the Bright/spectroscopic galaxy sample.
The top-panel show $\fomc$ increasing a Gaussian photo-z in the
Bright/spectroscopic sample, with four lines corresponding to FxB-All, F+B-All,
FxB-Counts and F+B-Counts. A vertical line at $\sigma_z = 0.0035(1+z)$ marks the
expected photo-z in a narrow-band photometric survey. The lower panel shows
the same-sky ratios and also including two ratios for the volume effect.}
\label{eff_photoz}
\xef

Fig. \ref{eff_photoz} (top panel) shows $\fomc$ for increasing redshift
uncertainties in the spectroscopic sample. Larger photo-z errors dramatically
decrease the performance.  The vertical line indicate the forecast for a
narrow-band survey, e.g. PAU \citeind{polpz}. For the fiducial redshift binning,
the narrow-band photo-z error provide comparable constraints to a fully
spectroscopic survey. This result depend on the redshift bin width. Thinner
bins are more sensitive to the photo-z value, with details provided in
paper-II. For larger errors of $\sigma_z = 0.05 (1+z)$, the information in the
galaxy clustering (Counts) drop by almost two orders of magnitude. This
decrease is likely overestimated because the number of bins is
fixed\footnote{We have \nbright\ bins and bias parameters for the spectroscopic sample
regardless of the photo-z uncertainties.}, but it demonstrates how the
2D-correlations in narrow bins benefit from the good redshift information.

The bottom panel (Fig. \ref{eff_photoz}) shows the same-sky benefit ratio
(FxB/F+B). When increasing the redshift uncertainties above $\sigma_{68}
\approx 0.015 (1+z)$, the ratios decline both for All and Counts. Including
lensing allows measuring counts-shear from either spectroscopic or photometric
foreground galaxies. This measurement depends on knowing the bias of the
foreground galaxies. Higher redshift uncertainties directly increase the
spectroscopic (B) bias error, while indirectly for the photometric (F) bias
from cross-correlation of galaxy counts in the two samples. For only galaxy
count, the weaker importance of Bright-Faint cross-correlations is partly
compensated by a loss in spectroscopic bias (in F+B), but the same-sky ratio
still declines. Two lines (\gkfb/F+B) show the direct benefit from overlapping
volume. The (\gkfb/F+B):All ratio decline fast, which means the sampling
variance cancellation is more dependent on good redshift resolution in the
spectroscopic sample.

\section{Conclusions}
In a series of articles (paper-I, II and III) we have looked at at combining the
information from photometric and spectroscopic surveys. The photometric surveys
measures galaxy shapes and can constrain cosmology from weak gravitational
lensing, while the accurate redshift determination in spectroscopic surveys is
suited for galaxy clustering, RSD and BAO. A central question is: Should
spectroscopic and photometric surveys ideally be over the same area? Previous
studies disagree, finding either none or very high same-sky benefit. This paper
summarized our understanding, building on the detailed study in paper-I, II and
III.

Section \ref{sec:assumpt} describe the forecast assumptions (for more details
see paper-II). The weak lensing and galaxy counts are both analyzed using
2D-correlations, with narrow bins in the spectroscopic sample. Radial
information of the spectroscopic sample is included through intrinsic
correlation between narrow redshift bins (see paper-I, paper-II,
\citedir{asorey}, \citedir{asorey2}). The difference between overlapping (FxB)
and non-overlapping (F+B) surveys is the additional cross-correlations and the
additional covariance since both surveys trace the same matter fluctuations. We
forecast the constraints using Fisher matrices, focusing on a combined figure
of merit ($\fomc$), which includes $w_0, w_a$ and the growth parameter
$\gamma$. Note that all assumptions exactly match paper-I, II and III. The
general ideas are the same as \citedir{gazta}, but the implementaion is quite
different.

The $\fomc$ is estimated for the photometric (F) and spectroscopic (B) surveys
alone and when combining them for overlapping (FxB) and non-overlapping (F+B)
surveys. We consider only including galaxy counts, the effect of removing
counts-shear cross-correlations and a special case to discuss the effect of
overlapping volumes. All those cases are estimated when fixing the bias or
ignoring the effect of WL, RSD or BAO. For the fiducial case (first row, first
column in Table \ref{tbl_fomc}), we find overlapping surveys benefit 50\% in
$\fomc$ and equivalently 30\% increase when only including galaxy counts.

One difference between overlapping and non-overlapping surveys are the
additional cross-correlations. In the overlapping surveys, one can
cross-correlate the foreground spectroscopic galaxy counts with the background
shear. Including counts-shear is possible using either foreground photometric
or spectroscopic galaxy counts. As presented in the main table, dropping either
set of cross-correlations give a small change, while dropping both dramatically
decrease the constraints. The benefit of additional correlations mainly comes from
counts-counts cross-correlations, which constrains bias and therefore make the
counts-shear more powerful. In \citedir{kirk} the authors acknowledge the strength
of counts-shear in the photometric (F) sample, but ignore them in the combined
forecast for technical reasons (private communication). This artificially
increase the same-sky (FxB/F+B) benefit and probably explain most of the
difference to our more modest same-sky benefit.

Overlapping surveys also directly improve constraints from the added
covariance. The galaxy over-densities in both surveys trace the same underlying
mass and the covariance leads to sample variance cancellations. Through a
special case (\gkfb) which can be though as either removing all
cross-correlations (between F and B) from FxB or adding covariance to the
non-overlapping (F+B) surveys, we find about equal benefit from sample variance
cancellation and the additional cross-correlations. In section
\ref{sec:example}, we demonstrate the non-intuitive effect of stronger
covariance giving better cosmological constraints using a simplified analytical
example. We show that for optimal constraints, one should minimize the
correlation (Pearson coefficient) between cosmological and nuisance parameters
and maximize the covariance between the nuisance parameters between the
spectroscopic and photometric surveys, which is achieved by increasing the
overlap between the two samples.

Previous analysis \mycite{mcdonald,deputter} combined a 3D power spectrum of
galaxy counts with 2D correlations for lensing, following \citedir{gazta}.
These analysis ignore the (radial) covariance between the 2D and 3D estimator.
Since the covariance between different tracers reduce the sample variance,
ignoring this covariance could partially explain their lower benefit of
overlapping surveys. These papers also ignore the significant radial 
information in the cross-correlation of the photometric and spectroscopic
survey (paper-I). The assumptions also differs, including this paper
using $l_{max}=300$ for both WL and Counts, while two papers combining
2D and 3D use $l_{max}=2000$ for WL and a larger $k_{max}$ for BAO, which
might affect the same-sky conclusion. Extending the forecast to non-linear
scales is left to future work.

Section \ref{sec:config} looked at the impact of galaxy density and redshift errors.
Starting from low galaxy densities in the B (spectroscopic) sample, increasing
the density strongly improves the $\fomc$. This benefit saturate around 0.4
gal/sq.arcmin. for all considered configurations. The shot-noise from low
densities decrease the benefit of sample variance cancellation. Increasing the
densities therefore rise the same-sky ratio and the biggest change occurs when
only galaxy counts are included. The last subsection investigated the
dependence on accurate redshift errors in the spectroscopic sample. Through
artificially increasing the spectroscopic redshift uncertainty, we find a
strong degradation in constraints as a function of  redshift accuracy.

In summary, this paper finds important gains from overlapping galaxy surveys.
The statistical benefit comes from both additional cross-correlations and
sample variance cancellations when using photometric and spectroscopic tracers.
We have studied the literature and believe we could plausibly explain the
discrepancies and confusion which is still surrounding the topic of overlapping
galaxy surveys. We also identify several effects that impact this result and
show an analytical example for the covariance from the overlapping surveys.  In
addition to reducing the statistical errors, the overlapping surveys can help
reducing systematic uncertainties, which is needed for the next generation of
galaxy surveys.

\section*{Acknowledgments}
We would like to thank the group within the DESI community looking at
overlapping surveys. M.E. wish to thank Ofer Lahav, Henk Hoekstra and Martin
Crocce in his thesis examination panel, where these results were discussed. E.G
would like to thank Pat McDonald for exploring and comparing results. Funding
for this project was partially provided by the Spanish Ministerio de Ciencia e
Innovacion (MICINN), project AYA2009-13936 and AYA2012-39559,
Consolider-Ingenio CSD2007- 00060, European Commission Marie Curie Initial
Training Network CosmoComp (PITN-GA-2009-238356) and research project 2009-
SGR-1398 from Generalitat de Catalunya. M.E. was supported by a FI grant from
Generalitat de Catalunya. M.E. also acknowledge support from the European
Research Council under FP7 grant number 279396. 

\bibliographystyle{mn2e}
\bibliography{sep.bib}{}
\end{document}